\documentclass[sigconf, nonacm=true]{acmart}
\usepackage{tabularx}
\usepackage{multirow}
\usepackage{hhline}
\usepackage{enumitem}
\usepackage{tcolorbox}

\copyrightyear{2019}
\acmYear{2019}
\acmConference[WCCCE '19]{Proceedings of the 24th Western Canadian Conference on Computing Education}{May 03--04, 2019}{Calgary, AB, Canada}
\acmPrice{0.00}
\begin{document}

%
% The "title" command has an optional parameter, allowing the author to define a "short title" to be used in page headers.
%\title{What are they talking about? Semi-supervised topic modeling for classifying discussion board activity}
\title{Tracing Forum Posts to MOOC Content using Topic Analysis} % takes up two lines :(
% \title{Tracing forum discussions to source material using topic analysis}

%
% The "author" command and its associated commands are used to define the authors and their affiliations.
% Of note is the shared affiliation of the first two authors, and the "authornote" and "authornotemark" commands
% used to denote shared contribution to the research.
\author{Alexander William Wong}
\email{alex.wong@ualberta.ca}
\orcid{0000-0002-9773-7484}
\affiliation{%
  \institution{University of Alberta}
  \streetaddress{259B2 Computing Science Centre}
  \city{Edmonton}
  \state{Alberta}
  \postcode{T6G 2S4}
}

\author{Ken Wong}
\email{kenw@cs.ualberta.ca}
\affiliation{%
  \institution{University of Alberta}
  \streetaddress{4-50 Athabasca Hall}
  \city{Edmonton}
  \state{Alberta}
  \postcode{T6G 2S4}
}

\author{Abram Hindle}
\email{abram.hindle@ualberta.ca}
\orcid{0000-0002-4373-4958}
\affiliation{%
  \institution{University of Alberta}
  \streetaddress{4-47 Athabasca Hall}
  \city{Edmonton}
  \state{Alberta}
  \postcode{T6G 2S4}
}

%
% By default, the full list of authors will be used in the page headers. Often, this list is too long, and will overlap
% other information printed in the page headers. This command allows the author to define a more concise list
% of authors' names for this purpose.
% \renewcommand{\shortauthors}{Wong, et al.}

% The abstract is a short summary of the work to be presented in the article.
\begin{abstract}
Massive Open Online Courses are educational programs that are open and accessible to a large number of people through the internet.
To facilitate learning, MOOC discussion forums exist where students and instructors communicate questions, answers, and thoughts related to the course.

The primary objective of this paper is to investigate tracing discussion forum posts back to course lecture videos and readings using topic analysis.
We utilize both unsupervised and supervised variants of Latent Dirichlet Allocation (LDA) to extract topics from course material and classify forum posts.
We validate our approach on posts bootstrapped from five Coursera courses and determine that topic models can be used to map student discussion posts back to the underlying course lecture or reading.
Labeled LDA outperforms unsupervised Hierarchical Dirichlet Process LDA and base LDA for our traceability task.
This research is useful as it provides an automated approach for clustering student discussions by course material, enabling instructors to quickly evaluate student misunderstanding of content and clarify materials accordingly.

\end{abstract}

% The code below is generated by the tool at http://dl.acm.org/ccs.cfm.


% Keywords. The author(s) should pick words that accurately describe the work being
% presented. Separate the keywords with commas.

% This command processes the author and affiliation and title information and builds
% the first part of the formatted document.
\maketitle

\section{Introduction and Motivation}
The primary feature distinguishing a Massive Open Online Course (MOOC) and a traditional course is the MOOC's capacity to scale to a seemingly limitless number of concurrent students~\cite{pappano2012year}.
The scalability of the MOOC can be represented by the ratio of instructors and students.
When students vastly outnumber the instructional staff, the opportunity for a student to interact meaningfully with the instructors diminishes~\cite{huang2014superposter}.
Furthermore, the instructional staff's ability to gauge student understanding is also hampered.
It is impractical for a fixed number of instructors to respond to the individual needs of an unbounded, growing number of students~\cite{mackness2010ideals}.

One method for students and instructors to interact is through the discussion forums.
This medium allows relevant conversation where students ask questions, express their thoughts, and seek help from their peers and instructors about the course material.
\citeauthor{stephens2014monitoring}'s research surveyed 92 MOOC instructors and determined that the conversations students have on course discussion forums are a useful repository of data for course self-evaluation~\cite{stephens2014monitoring}.
Unfortunately, basic visualizations showing discussion board summary statistics are not useful to instructors, as the rich semantic information about the discussions is lost~\cite{stephens2014monitoring}.

% AH: You know if you'd plot topic association over time of labelled LDA that might be
%     useful as a summary and contribution
% AH: worse, you claim to have this high level contribution but we don't see evidence of it

\begin{figure}
    \centering
    \includegraphics[width=\columnwidth]{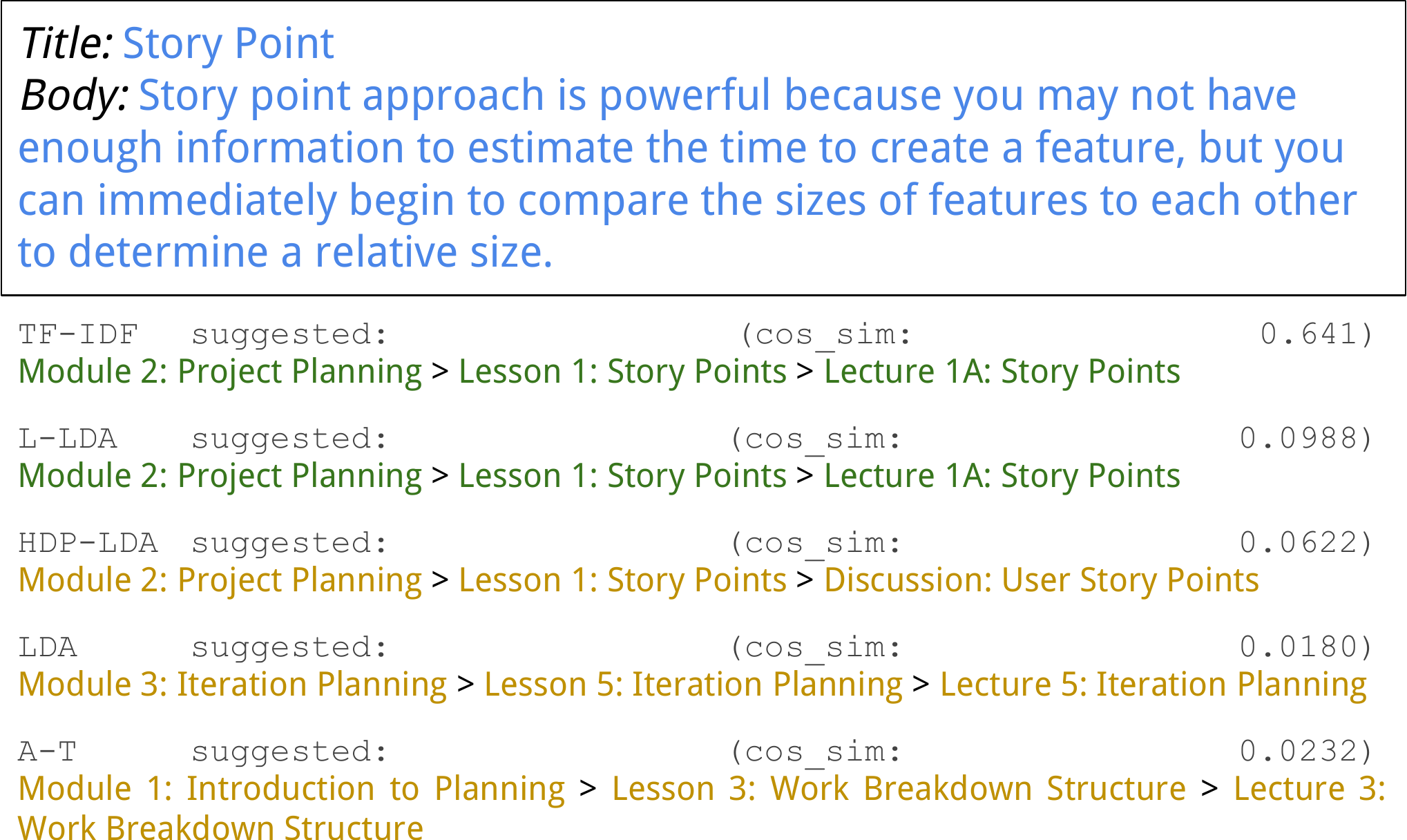}
    \caption{Sample discussion post from "Agile Planning for Software Products" classified by our topic models. TF-IDF and Labeled-LDA models matched researcher given label.}
    \label{fig:sample_classification}
\end{figure}

This paper presents a methodology to trace discussion forum activity back to the underlying course material, as demonstrated in Figure~\ref{fig:sample_classification}.
One motivation is to show instructors potential areas of misunderstanding with respect to a given lecture video or reading.
Instructors could view discussions grouped by lecture and ordered by post count.
This would allow instructors to more quickly respond to multiple students struggling with the same issue.
Our approach has the additional benefit that no additional labelling or overhead is required to perform this mapping.
We restrict ourselves to data that is readily available in the course material and discussion forum.

In this paper, we define a \textit{topic} as similar words that occur in a collection of text documents.
A topic model like LDA receives a set of input documents, a pre-defined number of topics $n$, and some additional set of prior parameters.
LDA then attempts to find a set of $n$ topics that describe the input documents.
A linear combination of all $n$ discovered word distributions can then be mapped to each input document.
Topics themselves can be thought of as a ranked list of words, ordered from high topic relevance to low topic relevance.

We utilize four different variants of Latent Dirichlet Allocation (LDA) to extract topic-document distribution feature vectors from discussion forum content and course material~\cite{blei2003latent}.
Term Frequency Inverse Document Frequency (TF-IDF), a normalization approach extracting number of times a word appears in the document divided by the total number of words in the document, is used as the baseline model for forum content feature extraction~\cite{SALTON1988513}.

We compare the effectiveness of unsupervised LDA approaches with labeled extensions of LDA for discussion traceability.
For labels, we utilize the defined course module, lesson, and item headings already associated with each course document.
Author~Topic~models train on text with associated authors, enabling association of extracted topics to the individuals that wrote the document~\cite{rosen2004author}.
Labeled~LDA trains on text with a set of tags, which enables inferring word association to the defined tags that exist within the input corpus~\cite{ramage2009labeled}.
Our study aims to answer the following research questions:

\begin{enumerate}[label=\textbf{RQ\arabic*:}]
    \item Are course materials an appropriate corpus for training discussion forum evaluating topic models? % maybe?
    \item Are topic extraction models useful in tracing discussion forum conversations back to MOOC content? % TF-IDF comparison
    \item Do supervised topic models outperform unsupervised topic models in tracing and classifying forum activity? % Compare MRR between supervised & unsupervised LDA models
\end{enumerate}

\section{Methodology}
We first relate MOOC discussion forum posts to topics extracted from MOOC course content.
Afterwards, we evaluate if topic mapping vectors are appropriate features to use for discussions and course material traceability.

Our methodology is to extract course material from the Coursera platform and convert the unstructured text into a natural language processing (NLP) ready format.
Then we perform topic analysis using unsupervised LDA, unsupervised HDP-LDA, supervised Author-Topic models, and Labeled LDA.
We perform topic model inference on discussion posts. A post is given weights indicating topic contributions to the post.
We compare our topic models to our baseline TF-IDF model to determine the utility of topic extraction models to create features from discussion posts and course material.
Finally, we present our results and analysis.

\subsection{Data Mining Coursera}
\begin{figure}
    \centering
    \includegraphics[width=\columnwidth]{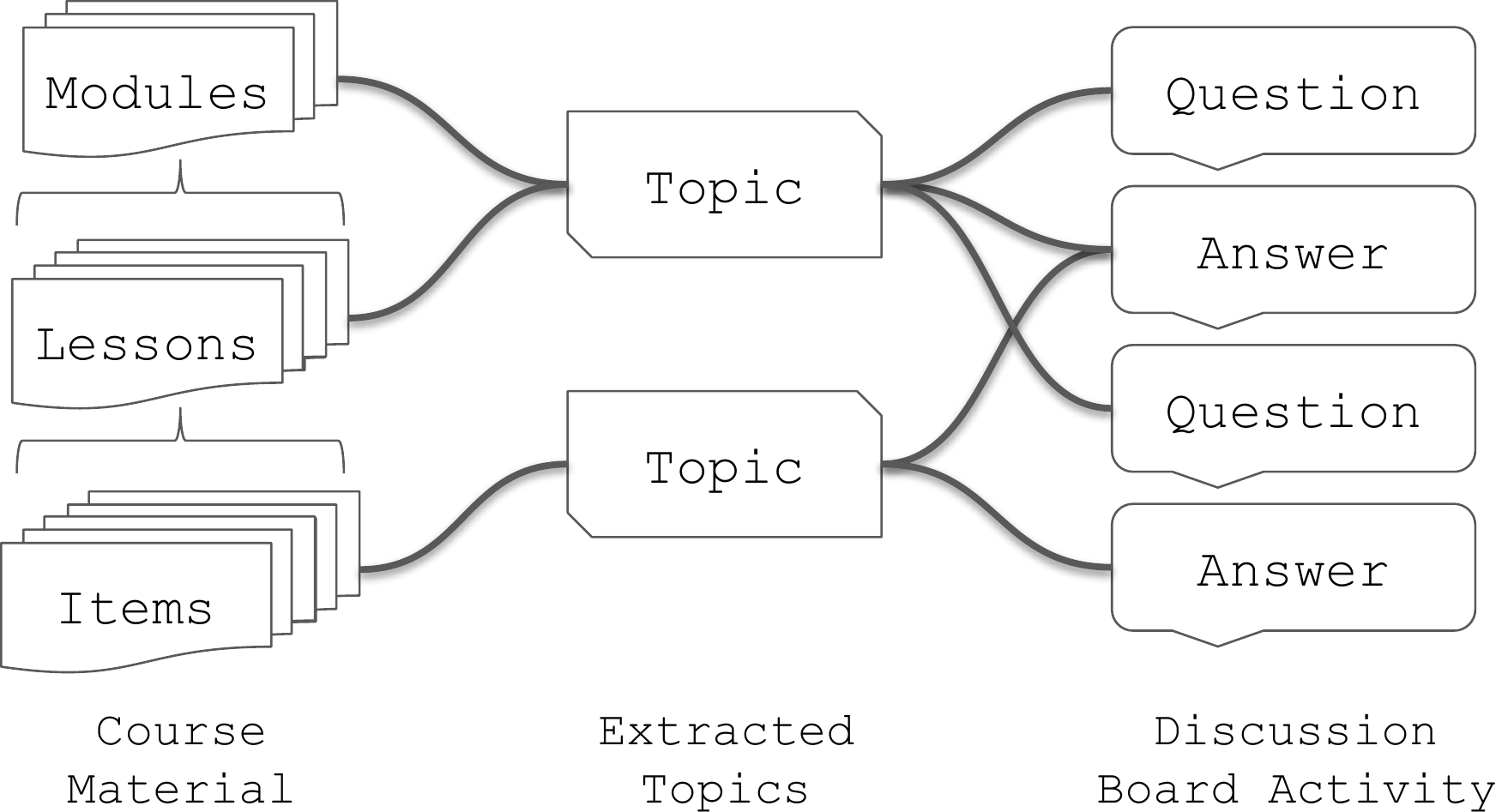}
    \caption{Course material consists of modules, lessons, and items. Topics infer discussion board activity. Topics have a many-to-many relationship to course material.}
    \label{fig:data_model}
\end{figure}

Coursera is an online learning platform that allows universities and other organizations to offer MOOCs, specializations, and degrees.
A MOOC can be versioned by \textit{branches} or be composed of a single \textit{branch}.
Figure~\ref{fig:data_model} shows the structure of course material within a \textit{branch}.
A \textit{branch} is composed of \textit{modules}, which are groups of \textit{lessons} intended to encompass a week's worth of material.
A \textit{lesson} is a more focused group containing \textit{items} on a specific subject matter.
An \textit{item} is the smallest document for Coursera MOOCs. \textit{Items} are used to encapsulate specific lecture videos, readings, quizzes, or assignments.

Multiple forums may exist for a single course.
The forums are curated by the instructional staff and exist to tailor discussion to a general domain.
Common course forum titles include "Introductions", "General Course Discussion", and "Technical Issues".
When interacting with the forums, students are limited to either posting \textit{Questions} or \textit{Answers}.
Questions are top level discussion entities that exist immediately underneath a forum, usually seeking subject matter clarification and help.
Answers are replies to Questions and typically contain hints and guidance for the related question.
In our study, we did not distinguish between the forums a user posted in or the type of discussion.

Only the most recent, active branch was evaluated for each course.
We extracted the hierarchical structure for each course, adding lecture video subtitles and readings to our available corpus.
Unstructured text from quizzes and assignments were ignored due to limitations arising from converting interactive student experiences into static documents.
All mined textual data was pre-processed by stripping XML tags (characters encapsulated by "\texttt{<}", and "\texttt{>}"), punctuation, consecutive white-spaces, numeric digits(\texttt{0} to \texttt{9}), stop words ("this", "and", "the", etc.), and words less than three characters in length.
The remaining words were then porter-stemmed to their root form (removal of "-ing", "-s", "-ed", "-ly", etc.).

The hierarchical structure of the course was obtained through privileged access to course material by course administrators.
The raw document information, such as lecture video subtitles and readings, was mined through polling the Coursera On-Demand API endpoint as an authenticated user enrolled in all of the relevant courses.
Figure \ref{fig:data_model} shows the scope of our data model for our research.

\begin{figure}
    \centering
    \includegraphics[width=\columnwidth]{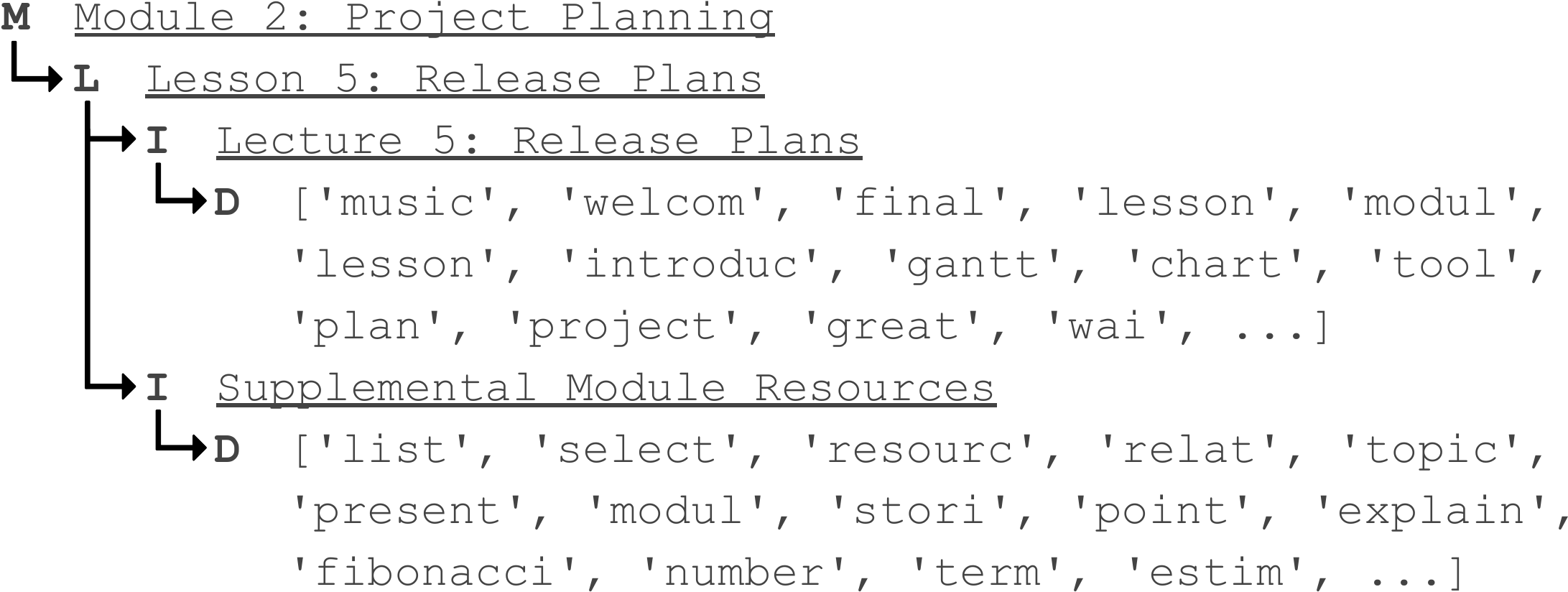}
    \caption{Two documents (D) from "Agile Planning for Software Products" with corresponding labels (M, L, I).}
    \label{fig:sample_doc}
\end{figure}

We performed supervised and unsupervised topic extraction on five
% University of Alberta
MOOCs run on the Coursera platform, described in Table~\ref{tab:uofa-courses}.
All courses studied were related to Computer Science and Software Engineering.
After pre-processing the course material, we obtained 190 documents, with 8,805 unique stemmed word stubs.
The average document contained 158 word stubs.
An example of the document structure and label hierarchy can be found in Figure~\ref{fig:sample_doc}.

\begin{table}
\begin{tabularx}{\columnwidth}{@{}p{0.52\columnwidth}ccc@{}}
    \toprule
    \multirow{2}{*}{Coursera Course Name} & Num. & Num. & Num. \\
    & Modules & Lessons & Items \\
    \midrule
    Agile Planning for Software Prod. & 4 & 20 & 38 \\
    Client Needs and Software Reqs. & 4 & 21 & 42 \\
    Design Patterns & 4 & 9 & 37 \\
    Introduction to SPM & 3 & 19 & 30 \\
    Object Oriented Design & 4 & 15 & 43 \\
    % Service-Oriented Architecture & 4 & 4 & 27 \\
    % Software Architecture & 4 & 13 & 30 \\
    \bottomrule
\end{tabularx}
\caption{Five analyzed Coursera courses and corresponding number of extracted modules, lessons, and items.}
\label{tab:uofa-courses}
\end{table}

\subsection{Topic Analysis Approaches}
Topics were extracted from the course content material using unsupervised and supervised LDA.
New models were trained for each course.

We used the \texttt{gensim@3.7.1} implementation of LDA, Hierarchical Dirichlet Process LDA (HDP-LDA) and Author-Topic model~\cite{rehurek_lrec}.
To form our unsupervised set of topics, we used both LDA and HDP-LDA.
All LDA parameters were set as the library provided defaults ($\text{num\_topics}=100$, $\alpha=0.01$, $\beta=0.001$).
HDP-LDA addresses one shortcoming of traditional LDA by using a Dirichlet process to capture the number of topics, rather than defining the number of topics a priori~\cite{wang2011online}.
We ran HDP-LDA with the default configuration provided by the library ($\kappa=1$, $\tau=64$, $K=15$, $T=150$, $\alpha=0.01$, $\gamma=1$, $\eta=0.01$).

The Author-Topic model, an extension of base LDA, was introduced by \citeauthor{rosen2004author} to correlate documents with authorship information to provide more details on the subject knowledge of the given author~\cite{rosen2004author}.
The Author component of the Author-Topic model is classically represented by labeling an existing corpus of documents with the author(s) of the document.
However, we modified the standard usage and replaced Author with \texttt{modules}, \texttt{lessons}, and \texttt{items}.
The underlying assumption was course subject knowledge labeled by subject headings could be analogous to course content labeled by the content author.
The parameter choices for the Author-Topic model was left as the library provided defaults ($\text{num\_topics}=100$, $\alpha=0,01$, $\beta=0.001$).

Additional topic analysis using Labeled LDA was run.
Unlike Author-Topic models, Labeled LDA outputs topics constrained to the labels defined in the corpus~\cite{ramage2009labeled}.
Classical applications of Labeled LDA are analyzing tagged blog entries, where a given blog may have multiple associated tags.
We extend an existing implementation of Labeled LDA found on GitHub, with slight modification for topic inference without training~\footnote{Labeled LDA in Python.
    \url{https://github.com/fann1993814/llda}
    %\url{https://github.com/shuyo/iir}
    %URL DOUBLE BLIND HIDDEN
    }.
Our parameter choices for Labeled LDA are $\alpha=0.01$, $\text{beta}=0.001$, with 50 iterations.
The differences between the four studied models are defined in Table~\ref{tab:topic_models}.
We use the \texttt{module}, \texttt{lesson}, and \texttt{item} names as labels for our Author-Topic and Labeled LDA approaches.
% AH: I don't understand where the labels are from. What are the labels?

\begin{table}
    \centering
    \begin{tabularx}{\columnwidth}{@{}llll@{}}
    \toprule
    Model & Labels & Feature Characteristic & Vector Size \\
    \midrule
    TF-IDF & \texttt{False} & \# unique words/all words & \# words \\
    LDA & \texttt{False} & topics: bounded, latent & set 100 \\
    HDP-LDA & \texttt{False} & topics: unbounded, latent  & cap 150 \\
    Author-Topic & \texttt{True} & topics: label distributed & set 100 \\
    Labeled LDA & \texttt{True} & topics: restricted to labels & \# labels \\
    \bottomrule
    \end{tabularx}
    \caption{Comparison of the baseline and four different topic analysis models used in our study.}
    \label{tab:topic_models}
\end{table}

\subsection{Relating Topics to Discussion Activity}
To derive a relationship between discussion forum activity and the course material, we used our trained topic models to infer topic distribution of the discussion form posts.
We inferred the relationship between the word distribution within the discussion post and the word distribution within the topic.
We applied the same pre-processing step from the course material on the discussion forum activity. We removed tags, punctuation, consecutive white-space, numbers, stop words, and words less than three characters in length before stemming all remaining words to their root word.

LDA inference takes as input a set of documents and estimates the topic weight distribution for each document.
Because LDA inference does not modify the existing model and no learning occurs, we performed inference using only the subset of words in the discussion entity that have appeared in our course material vocabulary.
No out of vocabulary tokens are used for evaluation.
These discussion questions and answers were related to pre-existing topics extracted from the course material.

\begin{equation}
    \text{Cosine Distance} = 1 - \frac {\pmb A \cdot \pmb B}{||\pmb A||_2 ||\pmb B||_2}
    \label{eq:cos_distance}
\end{equation}

If a discussion post has similar topics to a given lecture, we want to suggest that lecture as a likely candidate for the post.
Specifically, the document-topic vector of the discussion post is compared with the document-topic vector of the course item.
To determine topic similarity, we used cosine distance as defined in Equation~\ref{eq:cos_distance}.
The cosine distance function returns a floating value between 0 and 1.
As the cosine distance approaches 1, the two elements are more distinct.
The two elements are more similar as the cosine distance approaches 0.
We choose cosine distance as our measurement due to its success in prior literature in ranking topically-similar documents~\cite{6062077}.

% you're using cosine distance as a score for the best topic
% or the best document? What's the point?
% or are you going back to documents that the topic originated?
% what's the actual score here. The topic cosines or the document cosines or what

\subsection{Evaluating Discussion Classification}

% explain it better
% what is being returned? documents or topics
% how are they ranked: cosine distance
% are they in order: cosine distance

To determine the effectiveness of our models in tracing discussion forum text back to the underlying course content, we randomly sampled from the discussion activity and manually assigned the discussion to the underlying course item.
Posts were queried for manual labelling using \texttt{sqlite} "\texttt{ORDER BY RANDOM()}".
Only the primary author labeled the discussion posts.
Discussion posts were assigned a \texttt{module}, \texttt{lesson}, and \texttt{item} name.
Labelling the discussion posts was challenging as there were many discussions that did not have a clear mapping to one course item.
Discussion posts that were blatantly off topic were excluded from topic model traceability evaluation, using the researcher's judgment.
We evaluated our models on 100 manually classified posts for each of the 7 courses.

\begin{equation}
    \text{Mean Reciprocal Rank} = \frac{1}{|Q|} \sum^{|Q|}_{i=1}\frac{1}{\text{rank}_i}
    \label{eq:mrr}
\end{equation}

We evaluated each model-provided forum post ordering with our manually labeled discussion data using mean reciprocal rank (MRR), as shown in Equation~\ref{eq:mrr}.
Given a sample number of queries $Q$, we return the multiplicative inverse of the rank for the correct answer.
For example, if a model ranked a discussion question with our value in the first place, the MRR would be $1$.
Second and third place would be $\frac{1}{2}$ and $\frac{1}{3}$ respectively.

To evaluate performance we produced test sets (not training sets) by bootstrapping 1000 samples from 100 discussion-material document pairs that we had manually labeled for each of the five courses. 
We bootstrap in order to provide a better estimate of performance based on limited manual labelling and sampling.

% AH: I don't understand how this is qualitative. You're using MRRs...
% AH: is this some RQ?
\subsection{Qualitative Course Vocabulary Evaluation}
To help answer the first research question, we summarized the underlying themes of manually labeled off-topic discussions in the course context.
We looked at the vocabulary intersection of words used by students with words used in the lecture videos and readings.
Given common words used by students that did not appear in the course material, we used our judgment to explain the recurring off-topic themes that did not map cleanly back to course material.

\section{Results}
\begin{table*}[t]
    \centering
\noindent\begin{tabular}{r*{5}{|p{0.2\columnwidth}}|}
 \multicolumn{1}{c}{} & \multicolumn{1}{c}{TF-IDF} & \multicolumn{1}{c}{LDA} 
  & \multicolumn{1}{c}{HDP-LDA} & \multicolumn{1}{c}{Author-Topic} & \multicolumn{1}{c}{Labeled LDA} \\ \hhline{~*5{|-}|}
 Agile Planning for Software Products           & \bf 0.824 & 0.388 & 0.400 & 0.072 & \emph{0.436} \\ \hhline{~*5{|-}|}
 Client Needs and Software Requirements         & \bf 0.654 & 0.163 & \emph{0.478} & 0.362 & 0.246 \\ \hhline{~*5{|-}|} 
 Design Patterns                                & \bf 0.654 & 0.266 & 0.173 & 0.066 & \emph{0.390} \\ \hhline{~*5{|-}|}
 Introduction to Software Product Management    & \emph{0.187} & 0.077 & 0.172 & 0.050 & \bf 0.217 \\ \hhline{~*5{|-}|}
 Object Oriented Design                         & \bf 0.927 & 0.400 & 0.292 & 0.087 & \emph{0.432} \\ \hhline{~*5{|-}|} 
 \hhline{*6{|-}|}
 \textit{All Courses Combined}                  & \bf 0.649 & 0.259 & 0.303 & 0.127 & \emph{0.344} \\ \hhline{~*5{|-}|}
\end{tabular}\par\bigskip
    \caption{Bootstrapped mean reciprocal ranks for the baseline and topic models on our courses. Best values are bolded.}
    \label{tab:result_mrrs}
\end{table*}

\begin{figure}
    \centering
    \includegraphics[width=\columnwidth]{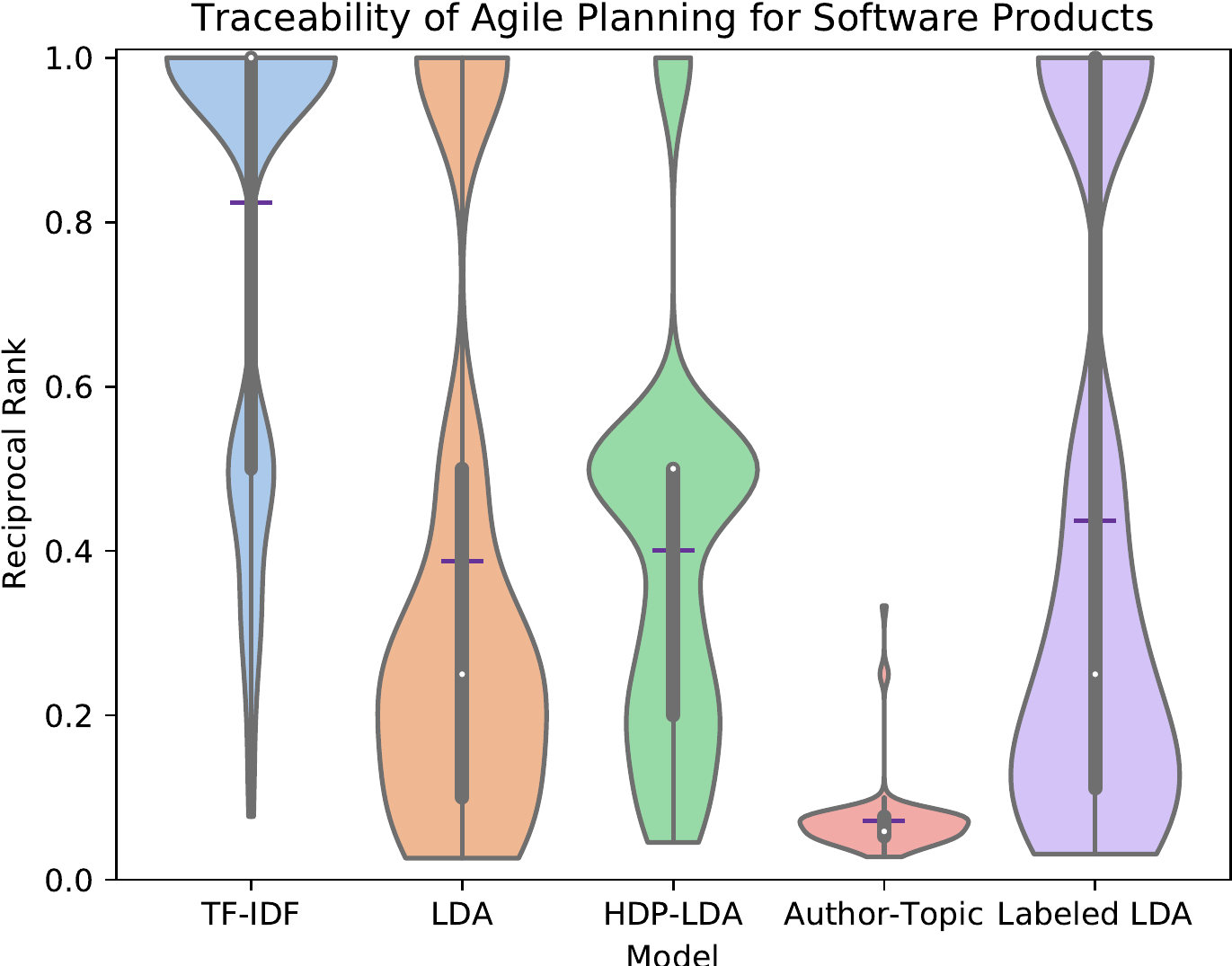}
    \caption{Reciprocal rank violin plot for "Agile Planning for Software Products". Horizontal line indicates model MRR.}
    \label{fig:agile-planning-plot}
\end{figure}

% AH: 1 thing about the results, the features used by TF-IDF were in the 1000s
% AH: the LDA results were at most 150?
The calculated MRRs derived from our analyzed courses are shown in Table~\ref{tab:result_mrrs}.
The baseline TF-IDF model outperforms topic models in all courses except for "Introduction to Software Product Management".
The least effective model for discussion forum traceability was the Author-Topic model, as it had the lowest MRR for four of the five courses.

A sample violin plot for one of our courses is shown in Figure~\ref{fig:agile-planning-plot}. %AH: this is not enough. Tell us what the plot means 
With the MRR closest to 1, the plot indicates that TF-IDF is the most accurate model for tracing discussion posts back to relevant course material.
Labeled LDA, HDP-LDA, and base LDA have MRRs around 0.4, meaning that the correct output is frequently ranked as the second or third place in the output.

\subsection{Course Vocabulary Applicability to Forums}
We looked at the frequently occurring words used by students that were not encompassed within our course vocabulary. Within "Agile Planning for Software Products", these common words included:

\begin{center}
\emph{
    peer, coursera, thank, gykcv, rntxi, submiss, ukr, submit, upload, kindli, gui, dev, lectur, regard, bug, request, happi, amazonaw, pdf
}
\end{center}

The five courses that were analyzed in our study contained peer assessments.
Other students were required to evaluate student submitted assignments.
The majority of discussion forum posts were requests to trade peer assessment grading.
These types of posts could still be mapped back to course material manually, as  assignment grading requests usually included plain text urls containing the associated module and lesson.

Other student discussion posts often included clarifications about lecture videos and readings, and applicability of course learned material on real world scenarios.
Many students expressed their fondness for the Agile methodology and gave concrete examples of their experiences with Agile after introducing the approach in their workplace and daily lives.
These two types of posts could also be mapped to relevant course lecture videos and readings, as there were clearly defined terminology used that gave appropriate context for their conversation.

Students also created many posts where they introduced themselves to the other members of the course. 
These types of posts were manually labelled as off topic, as no course material existed for individual student introductions.

\begin{tcolorbox}[sharp corners, top=1mm, bottom=1mm]
\textbf{RQ1}: For peer assessment requests and lecture video \& readings discussion, course material derived vocabulary was adequate for performing our traceability task.
\end{tcolorbox}

\subsection{Topic Model Utility in Forum Traceability}
We compared the MRRs of our models with random chance using the Wilcoxon Rank Sum Test with Continuity Correction.
Across all of the five courses, our trained topic models outperform random mapping with a $p\text{-value} < 2.2 * 10^{-16}$.

Topic models are less effective than the baseline TF-IDF model, however the correctly labeled output is still consistently within the top three ranks for the LDA, HDP-LDA, and Labelled LDA models. Author-Topic models performed the worst of all models, where the appropriate label was usually ranked ten places lower. % AH: are you making the claim because the MRR is 0.4?

Our result suggests that overall, topic models are a promising candidate for generating features from discussion forum activity which allow mapping back to the course material. % AH: one benefit is less features...

% AH: You really should do a TF-IDF + LDA model

\begin{tcolorbox}[sharp corners, top=1mm, bottom=1mm]
\textbf{RQ2}: Topic models outperform randomly mapping discussion posts to course material. This suggests that topic models are an effective candidate for traceability analysis.
\end{tcolorbox}

\subsection{Supervised vs. Unsupervised Topic Models}

To validate the effectiveness of supervised versus unsupervised topic models, we compared the Labeled LDA MRRs with the LDA and HDP-LDA models.
We ignored comparison tests using Author-Topic models due to their relatively poor performance.

We compared Labelled LDA with base LDA and HDP-LDA for each course using Wilcoxon Rank Sum Test with Continuity Correction.
All calculated $p$-values are displayed in Table~\ref{tab:sup-unsup-test}.
This table shows that the means for our three models are statistically different from each other.

Labelled LDA provided statistically significant higher mean reciprocal ranks compared with HDP-LDA and base LDA.
These results suggest that the use of labels for topic model analysis is important for traceability.
Although supervised topic models do not always outperform unsupervised approaches, labelled topic models appear to better infer traceability features for student discussion posts.

\begin{tcolorbox}[sharp corners, top=1mm, bottom=1mm]
\textbf{RQ3}: Properly labeled, supervised topic models perform better than unsupervised topic models in tracing discussion posts back to MOOC content.
\end{tcolorbox}
% not bootstrapped
% \begin{table}
%     \centering
%     \begin{tabular}{l*{2}{c}}
%     \toprule
%     Coursera Course Name & HDP-LDA & LDA \\
%     \midrule
%     Agile Planning for Software Prod. & 0.5176 & 0.2522 \\
%     Client Needs and Software Reqs. & $\mathbf{3.003*10^{-4}}$ & 0.1196 \\
%     Design Patterns & 0.3606 & 0.187 \\
%     Introduction to SPM & 0.06331 & $\mathbf{2.397*10^{-16}}$ \\
%     Object Oriented Design & 0.04447 & 0.1891 \\
%     % \textit{All Courses Combined} & 0.6274 & \bf 6.689e-4 \\
%     \bottomrule
%     \end{tabular}
%     \caption{Wilcoxon rank sum test generated $p$-values for Labelled LDA against HDP-LDA and LDA. Bold values reject the null hypothesis ($\alpha=0.01$).}
%     \label{tab:sup-unsup-test}
% \end{table}

\begin{table}
    \centering
    \begin{tabular}{l*{2}{c}}
    \toprule
    Comparison & W & $p\text{-value}$ \\
    \midrule
    Labeled LDA, LDA & 6478600 & $2.2*10^{-16}$ \\
    Labeled LDA, HDP-LDA & 5589900 & $4.322*10^{-9}$ \\
    \bottomrule
    \end{tabular}
    \caption{Wilcoxon rank sum test results for Labelled LDA against HDP-LDA and LDA using combined course data.}
    \label{tab:sup-unsup-test}
\end{table}

\section{Discussion}
Topic models have shown promise in mapping student discussion posts back to the underlying course material.
Compared to TF-IDF, our study showed that topic models could not achieve the same level of performance.
However topic models are advantageous in that their outputted vectors have constant size as defined by the number of topics used.
TF-IDF models output vectors that have a length upper bound equal to the size of the entire vocabulary.
It is more efficient to use topic models for documents with large vocabularies, as computing the document vector and performing the cosine similarity comparison is much slower using TF-IDF.

We did not attempt to label any of the model extracted topics directly, as they were only proxies for our traceability evaluation.
Additionally, we did not train a comprehensive model on the entire set of course material and discussion forum documents, as we wanted comparable baselines between our labelled and unlabelled topic models.

\section{Threats to Validity}
We acknowledge \textit{construct validity} threats, namely we used courses from a single institution that discussed computer science and software engineering material. We mitigated this threat by using multiple courses rather than one single course.
The primary threat to our research validity was the fact that only a single author labeled the evaluation dataset, therefore evaluation relied on a single author's judgment.
Potential bias could have been mitigated with multiple individuals labelling the dataset and reporting inter-rater reliability.

\section{Related Work}
This paper contributes to research in software engineering traceability and discussion forum topic analysis.

\subsection{Traceability \& Topic Analysis}
Tracing discussion forum content back to course material has similarities to the problem of tracing source code back to the specified requirements.
Prior work done by \citeauthor{4556122} showed that term frequency-inverse document frequency (TF-IDF) models can encode documents into a rich vector space model, allowing for specification traceability across software artifacts~\cite{4556122}.
\citeauthor{Tata:2007:EST:1328854.1328855} presented an approach for using TF-IDF models to compare two documents through cosine similarity~\cite{Tata:2007:EST:1328854.1328855}.
% \citeauthor{895989} outlined a framework for tracing stakeholder requirements through to development processes output, like source code~\cite{895989}.
\citeauthor{hindle2012relating}'s research showed that topics extracted from LDA on software requirements documentation can be traced to corresponding version control commits~\cite{hindle2012relating}.
Our research extends the state of the art in software engineering traceability by showing empirical results using labeled topic models for the same task.

\subsection{Discussion Forum Analysis}
The analysis of discussion forums in an educational context is a well-studied domain, spanning insight from sentiment analysis, social network interactions, user reputation, content popularity, and forum usage semantics~\cite{wen2014sentiment,wong2015analysis,coetzee2014should,breslow2013studying,onah2014exploring}.

Of prior work utilizing topic analysis, \citeauthor{ezen2015unsupervised} explored an unsupervised approach for clustering MOOC discussion board activity. Their strategy involved using the $k$-medoids clustering algorithm on bag-of-words inputs derived from discussion forum entities, then applying LDA to extract topics that the researchers could assign instructor meaningful labels to~\cite{ezen2015unsupervised}.
\citeauthor{atapattu2016topic} used LDA to capture influential topic clusters and isolate discussions requiring intervention on Coursera MOOCs~\cite{atapattu2016topic}.
\citeauthor{ramesh2014understanding}'s research predicted MOOC student survival using features extracted from seeded topic models on discussion forum posts~\cite{ramesh2014understanding}.
Topic analysis was performed by \citeauthor{Chen:2016:TME:2883851.2883951} on students' reflection journals, showing exploration of underlying themes and enabling prediction of journal grades~\cite{Chen:2016:TME:2883851.2883951}.
% \citeauthor{WISE201711}'s research proposed a linguistic natural language model for classifying discussion threads based on the relation to the course content and found reasonable predictive ability~\cite{WISE201711}.

\section{Future Work}
One limitation of our current approach is the inability to handle off topic discussions.
Many discussions were either unrelated to the course, or were too general to be mapped appropriately to a single course item.
Future relevant work includes investigating how to best model "Off-Topic" discussion posts, and how to accommodate course material relevance granularity.

We provide the source code used to perform this experiment in the intent that replication studies of this work can be performed on other MOOCs~\footnote{Source code for research experiment.
    \url{https://github.com/awwong1/topic-traceability}
    %URL DOUBLE BLIND HIDDEN
    }.
Our study focuses entirely on computer science and software engineering courses.
Another potential replication study could measure if discussion topicality matches course material across courses in Arts and Humanities, Business, Mathematics, Science, and other varying domains.

\section{Conclusion}
We performed an evaluation of using multiple variants of LDA topic models to trace discussion forum posts back to its corresponding course material in five MOOCs.
We investigated the traceability accuracy and found that although topic models could not surpass the TF-IDF model baseline, our trained topic models were significantly better than random chance.

We have provided results demonstrating that topic models can be used as a feature extraction step for unstructured text traceability.
These topic models do not require any additional data that does not already exist in MOOCs.
Unlike TF-IDF, they are fast and generate constant size feature vectors for similarity comparisons.
Our research can benefit MOOC stakeholders and advance other domains dealing with traceability.

\bibliographystyle{ACM-Reference-Format}
\bibliography{main}

\end{document}